\documentclass[sigconf,screen]{acmart}
\AtBeginDocument{%
  \providecommand\BibTeX{{%
    \normalfont B\kern-0.5em{\scshape i\kern-0.25em b}\kern-0.8em\TeX}}}

\usepackage{array}
\usepackage{subfigure}
\usepackage{graphicx}
\usepackage[switch]{lineno}
\usepackage{amsmath}
\usepackage{multirow}
\usepackage{booktabs}
\usepackage{bbding}
\usepackage{algorithm}
\usepackage{algorithmic}
\usepackage{xcolor}
\usepackage{xspace}
\usepackage{arydshln}
\usepackage{multicol}

\newcommand{\paratitle}[1]{\vspace{1.5ex}\noindent\textbf{#1}}
\newcommand{\ie}{\emph{i.e.,}\xspace}
\newcommand{\eg}{\emph{e.g.,}\xspace}
\newcommand{\baby}{\textsc{MACL}\xspace}
\newcommand{\babyx}{\textsc{MACL}}
\newcommand{\fig}{Figure\xspace}

\copyrightyear{2025}
\acmYear{2025}
\acmConference[arXiv]{Make sure to enter the correct
  conference title from your rights confirmation email}{June, 2025}

\begin{document}

\title{Rethinking Contrastive Learning in Session-based Recommendation}

\author{Xiaokun Zhang}
\affiliation{%
  \institution{Dalian University of Technology}
  \city{Dalian}
  \country{China}
}
\email{dawnkun1993@gmail.com}

\author{Bo Xu}
\affiliation{%
  \institution{Dalian University of Technology}
  \city{Dalian}
  \country{China}
  }
\email{xubo@dlut.edu.cn}

\author{Fenglong Ma}
\affiliation{%
  \institution{Pennsylvania State University}
  \city{State College}
  \country{United States}
  }
\email{fenglong@psu.edu}

\author{Zhizheng Wang}
\affiliation{%
  \institution{Dalian University of Technology}
  \city{Dalian}
  \country{China}
  }
\email{wzz_dllg@mail.dlut.edu.en}

\author{Liang Yang}
\affiliation{%
  \institution{Dalian University of Technology}
  \city{Dalian}
  \country{China}
  }
\email{liang@dlut.edu.cn}

\author{Hongfei Lin}
\authornote{Corresponding Author.}
\affiliation{%
  \institution{Dalian University of Technology}
  \city{Dalian}
  \country{China}
  }
\email{hflin@dlut.edu.cn}

\renewcommand{\shortauthors}{Xiaokun Zhang, et al.}

\begin{abstract}
Session-based recommendation aims to predict intents of anonymous users based on limited behaviors. With the ability in alleviating data sparsity, contrastive learning is prevailing in the task. However, we spot that existing contrastive learning based methods still suffer from three obstacles: (1) they overlook item-level sparsity and primarily focus on session-level sparsity; (2) they typically augment sessions using item IDs like crop, mask and reorder, failing to ensure the semantic consistency of augmented views; (3) they treat all positive-negative signals equally, without considering their varying utility. To this end, we propose a novel multi-modal adaptive contrastive learning framework called \baby for session-based recommendation. In \baby, a multi-modal augmentation is devised to generate semantically consistent views at both item and session levels by leveraging item multi-modal features. Besides, we present an adaptive contrastive loss that distinguishes varying contributions of positive-negative signals to improve self-supervised learning. Extensive experiments on three real-world datasets demonstrate the superiority of \baby over state-of-the-art methods. 
\end{abstract}




\maketitle

\section{Introduction}

Session-based Recommendation (SBR) focuses on predicting the intents of anonymous users based on limited behavioral data~\cite{GRU4Rec, DIMO}. Unlike traditional recommender systems that require users' profiles and long-term behaviors, SBR can provide personalized services based solely on limited behaviors observed within a short period, typically referred to as a session. This characteristic makes SBR suitable for practical scenarios where long-term user behaviors are either unavailable due to privacy policies or simply not recorded. Owing to its significant practical value, SBR has gained considerable attention in both academia and industry~\cite{NARM, Li@WSDM2023}.
\begin{figure}[t]
\centering
    \includegraphics[width=0.99\linewidth]{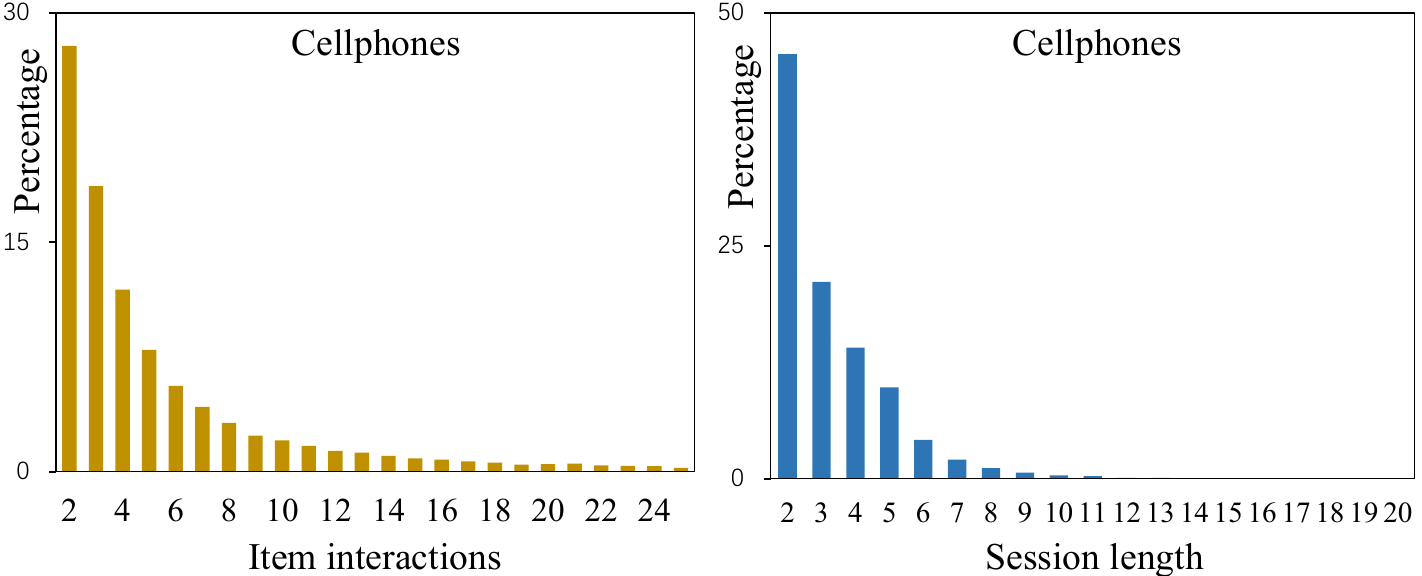}
    \caption{Data sparsity existing in SBR: (a) item-level long-tail items; (b) session-level short sessions.}\label{datasparsity}
\end{figure}
In the mean time, SBR encounters intractable challenges due to inherent data sparsity issues~\cite{DHCN,Yang@SIGIR2023,Kim@SIGIR2023}. Firstly, SBR is confronted with the problem of \emph{long-tail items} (\ie \textbf{item-level sparsity}). As shown in~\fig~\ref{datasparsity} (a), with a vast number of items available, users tend to interact with only a small fraction of them, resulting in a large number of long-tail items with extremely sparse interactions. The sparsity of long-tail items makes it challenging to accurately access user interest, consequently leading to a decline in recommendation performance~\cite{Yang@SIGIR2023}. Secondly, SBR is susceptible to \emph{short sessions} (\ie \textbf{session-level sparsity}). As illustrated in~\fig~\ref{datasparsity} (b), a typical session consists of only a few items, often fewer than five. Clearly, inferring user intents from such limited information is nontrivial~\cite{Xie@ICDE2022, DIMO}. These data sparsity issues serve as major bottlenecks that hinder the effectiveness of SBR in practice. 

Contrastive Learning (CL), which involves data augmentation to enrich data and uses a contrastive loss for self-supervised learning, has shown promising results in mitigating data sparsity ~\cite{chen@ICML2020}. 
Consequently, recent methods commonly employ CL to address the issue in SBR~\cite{Xie@ICDE2022}. 
These approaches typically apply random operations, such as crop, mask and reorder, to augment sessions based on item IDs. They assume that the augmented views retain similar semantics, allowing them to leverage CL for session embeddings enhancement by bringing positives together and pushing negatives apart. 
Through data enrichment and self-supervised learning, CL-based methods achieve improvements in SBR.

However, existing solutions of contrastive learning in SBR still suffer from following three significant obstacles:

\textbf{Gap 1} There is a lack of explicit efforts to address \textbf{item-level sparsity}. Current CL-based methods primarily focus on augmenting sessions via disturbing item ID sequences, but they overlook the issue of item-level sparsity. In fact, an individual item ID is hard to augment, leading to the failure of current methods in performing \emph{data augmentation at the item level}. By disregarding long-tail items, these methods fail to achieve satisfactory performance in practical scenarios.

\textbf{Gap 2} For \textbf{session-level sparsity}, augmentation methods of CL-based models predominantly rely on item IDs with random operations. However, item IDs merely serve as symbolic identifications and do not capture item specific semantics, like style and color. Consequently, relying on augmenting item IDs alone, these methods fail to ensure the \emph{semantic consistency of augmented views}, thereby resulting in limited effectiveness in session-level augmentation.

\textbf{Gap 3} In contrastive learning, distinct positive-negative pairs exhibit \textbf{varying utility} in terms of their impact on model performance~\cite{Khosla@NIPS2020}. Current methods overlook the differentiation between informative and uninformative signals, which significantly hampers effective self-supervised learning.

To address these issues, we propose a novel framework \underline{M}ulti-modal \underline{A}daptive \underline{C}ontrastive \underline{L}earning (\baby) for session-based recommendation. 
In \baby, a \textbf{multi-modal augmentation} is devised to achieve semantically consistent augmentations at both item and session levels. Unlike underrepresented item IDs, item multi-modal features (\eg images and text) contain rich semantics and can be augmented using various techniques from the computer vision (CV) and natural language processing (NLP) domains. As a result, we incorporate multi-modal features into both item- and session-level augmentations. For item-level augmentation, instead of relying on item IDs, we use item multi-modal features as additional signals for data augmentation. Specifically, we employ augmentation techniques from the CV and NLP to process an item's multi-modal features, achieving item-level augmentation (\textbf{Gap 1}). The selected augmentation techniques are designed to preserve the item’s original semantics during the augmentation, guaranteeing the semantic consistency of data augmentation. For session-level augmentation, we apply the augmented technique to all items within a session, forming an augmented session. Since the augmentation techniques maintain item semantics, the augmented session and its anchor session share consistent semantics, thereby ensuring semantically consistent augmentations at the session level(\textbf{Gap 2}). 
In addition, we present an \textbf{adaptive contrastive loss} to handle the varying utility of positive-negative signals (\textbf{Gap 3}). Unlike existing methods that assign the same weight to all positive-negative signals~\cite{Xie@ICDE2022, Du@CIKM2022}, our proposed adaptive contrastive loss evaluates the contribution of each signal on model performance using a simple yet effective neural network. Accordingly, this adaptive approach automatically highlights informative signals and diminishes uninformative ones during training, thereby facilitating effective self-supervised learning. Ultimately, we integrate user intent prediction and self-supervised tasks to enhance \baby. 
In summary, the main contributions of our work are listed as follows:

    
    
    
\begin{itemize}
    \item We scrutinize CL-based methods in SBR and identify their limitations including inability to handle item-level sparsity, corrupt data augmentation and failure in distinguishing varying utility of contrastive signals.
    
    \item We propose a novel \baby to overcome the above issues for SBR. In \baby, a multi-modal augmentation is devised to generate semantically consistent augmented views for both items and sessions. Besides, an adaptive contrastive loss is presented to reformulate the contrastive loss function by exploring the varying utility of contrastive signals. 
    
    \item Extensive experiments conducted on three real-world datasets demonstrate the superiority of \baby over state-of-the-art baselines. Further analysis also confirms the effectiveness of \baby in tackling data sparsity issues of SBR.
\end{itemize}

\section{Revisiting CL in SBR}\label{sec:revisit}
Contrastive Learning (CL) has been a potent antidote to session-level sparsity in SBR~\cite{Xie@ICDE2022, Du@CIKM2022}.
In this part, we revisit existing CL-based methods in SBR from their two key components: data augmentation and contrastive loss.

\subsection{Data Augmentation}\label{sec:dataaug}


Data augmentation plays a crucial role in CL since the models rely on it to enrich data. 
Existing operations are listed as follows:

\textbf{Crop} randomly crops continuous items in a sequence~\cite{Xie@ICDE2022, Zhou@CIKM2020}.

\textbf{Mask} randomly masks a portion of items in a sequence (without considering item order)~\cite{Xie@ICDE2022}.

\textbf{Reorder} randomly shuffles some items in a sequence ~\cite{Xie@ICDE2022, Du@CIKM2022}.

\textbf{Substitute} or \textbf{Insert} randomly replaces or inserts some correlated items in a sequence. The correlation between items is determined by co-occurrence frequency~\cite{Liu@CoRR}.

\textbf{Dropout} inputs embeddings into distinct dropout layers and outputs are assumed to be semantically similar~\cite{Qiu@WSDM2022}.

\textbf{Retrieval} views sessions sharing same label as positives~\cite{Qiu@WSDM2022, LXW@WSDM2023}. 

\textbf{Relation Mapping} examines data relations from different perspectives, and embeddings learned from these perspectives are viewed as augmented views~\cite{DHCN, Xia@CIKM2021, WZY@CIKM2022}.

\begin{figure}[t]
  \centering
  \includegraphics[width=0.95\linewidth]{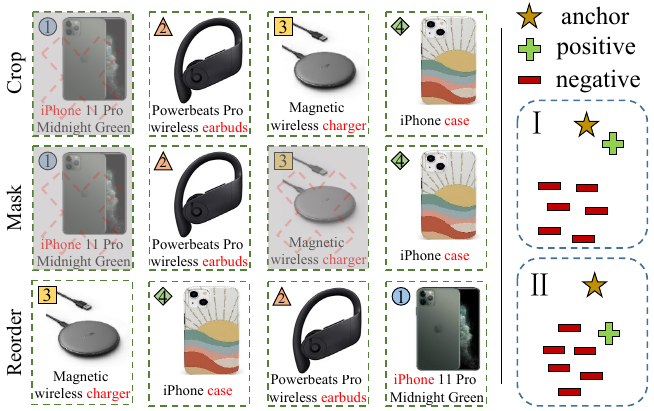}
  \caption{(a) Corrupt data augmentation operations; (b) Various utility of contrastive signals. }\label{augmentation}
\end{figure}

Existing methods, while attempting to enrich the data, do not guarantee semantically consistent augmentations. As depicted in~\fig~\ref{augmentation}(a), commonly used augmentation techniques can easily corrupt session semantics. For example, Crop and Mask operations may remove crucial items (like the iPhone), which can hinder the recommendation of related items (\ie accessories). 
Reorder can distort sequential patterns, as users often purchase cellphones before accessories. Substitute and Insert may introduce noise as they rely on weak item correlations. Dropout, duo to its high randomness, does not ensure semantic similarity either. 
Retrieval-based augmentation has a rigid rule, which limits its ability to generate augmented views for certain instances. Relation Mapping requires extensive knowledge to pre-define relations, which severely restricts its applicability. 
To address these limitations, we propose a multi-modal augmentation that leverages item multi-modal features with rich semantics to conduct robust data augmentation.

\subsection{Contrastive Loss}
Given an anchor $\mathbf{r}$, its positive $\mathbf{r}^{+}$ and negatives $\mathbf{r}^{-}$ = \{$\mathbf{r}_{1}^{-}$, $\mathbf{r}_{2}^{-}$, $\cdots$, $\mathbf{r}_{N}^{-}$\}, CL-based methods~\cite{Xie@ICDE2022, Liu@CoRR, LXW@WSDM2023, Xia@CIKM2021} improve representations by bringing positives close while pushing negatives apart as follows:
\begin{align}
        \mathcal{L}_{con} = -\sum \mathcal{L}_{r}  = -\sum \frac{<\mathbf{r}, \mathbf{r}^{+}>}{\sum_{k}<\mathbf{r},  \mathbf{r}_{k}^{-}>} ,
\end{align}
where $<,>$ is similarity function. In Eq.(1), every positive-negative signal \{$\mathbf{r}$, $\mathbf{r}^+$, $\mathbf{r}^-$\} is assumed to contribute equally for CL. However, as depicted in \fig ~\ref{augmentation} (b), distinct positive-negative signals possess distinct quality, as Set \uppercase\expandafter{\romannumeral1} apparently is better than Set \uppercase\expandafter{\romannumeral2}. In this work, we propose an adaptive contrastive loss to reweight contrastive signals by evaluating their utility to model performance for effective self-supervised learning.

\section{Preliminaries}

\subsection{Problem Formulation}
Let $x_i \in \mathcal{I}$ denote an item, where $\mathcal{I}$ is the unique item set and $|\mathcal{I}| = n$ is the number of items. Normally, a user evaluates an item based on its multi-modal features displayed on the website, including item image $x_i^{img}$ and description text $x_i^{txt}$. Notably, these features contain rich semantic information about an item, such as style, color, and size. The $\mathcal{S}$ = [$x_1, x_2, ..., x_m$] signifies a session produced by an anonymous user within a short period, where $x_i$ $\in$ $\mathcal{I}$ and $m$ is the session length. SBR aims to predict the next interacted item $x_{m+1}$ based on $\mathcal{S}$. 


\subsection{Representation Initialization}
For an item $x_i$, we can portray it from multiple perspectives like item ID $x_i^{ID}$ indicating item co-occurrence relations, item image $x_i^{img}$ and text $x_i^{txt}$ signifying its semantic information. Obviously, we should initialize such various information to serve as inputs for neural networks.

\paratitle{ID embedding.} 
Similar as conventional methods~\cite{NARM, BERT4Rec, SR-GNN}, we obtain an item ID embedding $\mathbf{e}_i^{id} \in \mathbb{R}^d$ associated with its ID $x_i^{ID}$ based on a look-up embedding layer ${\rm idEmb}(\cdot)$ via,
\begin{align}
    \mathbf{e}_i^{id} &= {\rm idEmb}(x_i^{ID}).
\end{align}

\paratitle{Image embedding.} 
The GoogLeNet~\cite{googlenet} is effective in extracting semantics from images. Thus, we apply it to obtain image embedding $\mathbf{e}_i^{img} \in \mathbb{R}^d$ from item image $x_i^{img}$ via,
\begin{align}
    \mathbf{e}_i^{img} &= {\rm imgEmb}(x_i^{img}),
\end{align}
where ${\rm imgEmb}(\cdot)$ is the GoogLeNet model pre-trained on a large number of images. 

\paratitle{Text embedding.} 
Due to its powerful ability to encode text, BERT~\cite{BERT} is employed to learn text embedding $\mathbf{e}_i^{txt} \in \mathbb{R}^d$ from item description text $x_i^{txt}$ via,
\begin{align}
    \mathbf{e}_i^{txt} &= {\rm textEmb}(x_i^{txt}),
\end{align}
where ${\rm textEmb}(\cdot)$ denotes the BERT model pre-trained on large text corpus.

\section{Methodology}
\begin{figure*}[t]
  \centering
  \includegraphics[width=0.90\linewidth]{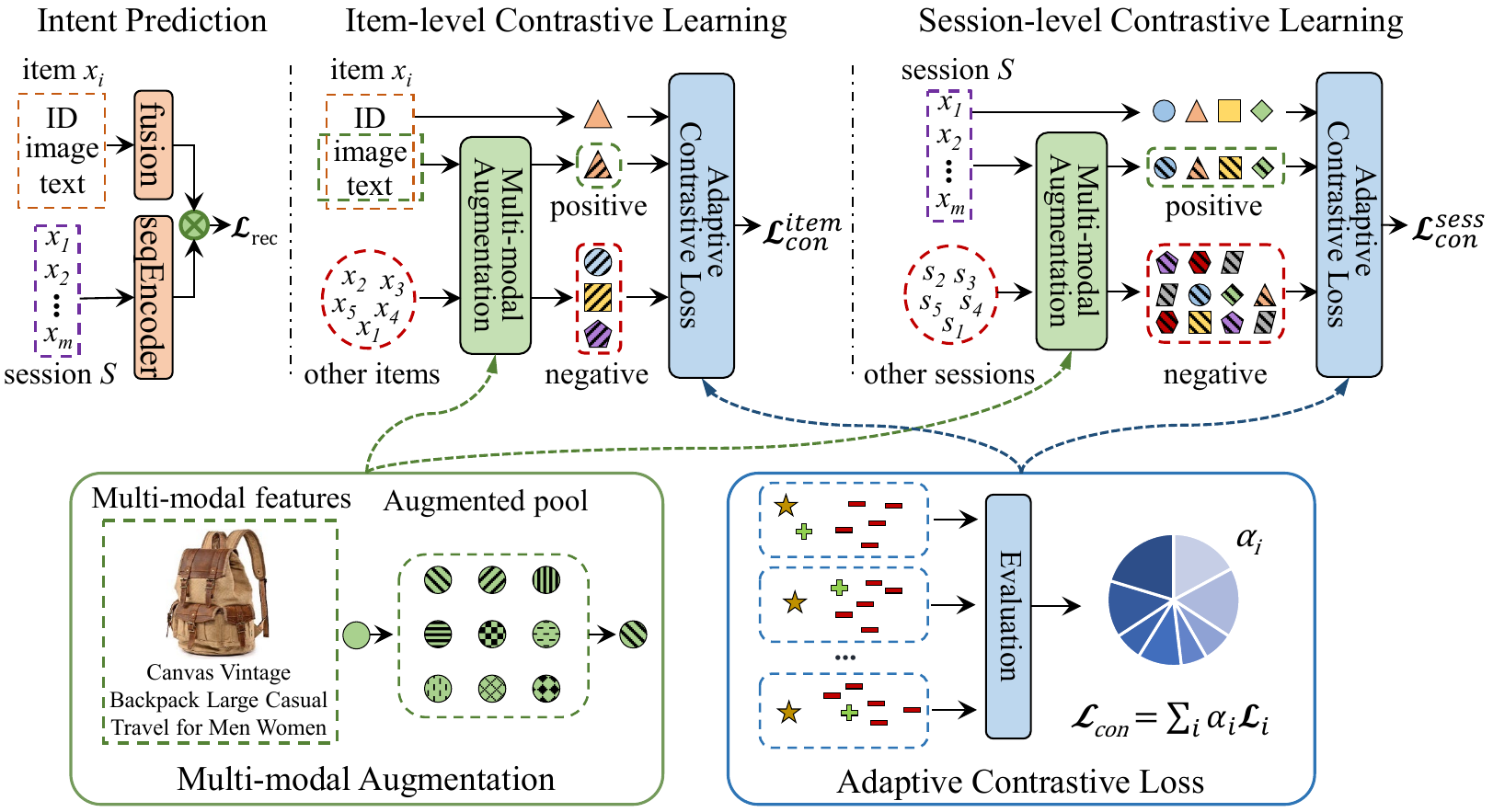}
  \caption{The graphical architecture of \baby. Intent prediction provides recommendations based on item and session embeddings. We solve item-level and session-level sparsity via item-level and session-level contrastive learning respectively. A multi-modal augmentation is devised to obtain semantically consistent augmented views based on item multi-modal features for both item and session augmentation. An adaptive contrastive loss is then presented to reformulate item and session-level contrastive loss via evaluating the varying utility of contrastive signals.(\textit{Best viewed in color})}\label{macl}
\end{figure*}

In this section, we will elaborate on the proposed \baby which is illustrated in \fig~\ref{macl}. We first formulate the recommendation task based on item and session embedding learning in the \emph{intent prediction}. The \emph{multi-modal augmentation} is then devised to obtain augmented views with consistent semantics at both item and session levels by leveraging item multi-modal features. Afterward, we present an \emph{adaptive contrastive loss} to distinguish informative signals from uninformative ones for effective contrastive learning. Finally, we improve \baby by jointly optimizing intent prediction and self-supervised learning in the \emph{training and inferring}.

\subsection{Intent Prediction}

SBR aims to reveal the intents of an anonymous user based on her limited behaviors within a session. With powerful representation ability, prevailing neural models for SBR focus on learning item and session embeddings and formulate a recommendation list by evaluating their similarity. In this part, we will detail the learning of item and session embedding and the formulation of the recommendation list in \baby.

\subsubsection{Item embedding learning}
Normally, a user decides whether to interact with an item after evaluating its multi-modal features displayed on the website like item image and description text~\cite{MMSBR,zhang2024side,SIDSBR}. Thus, besides item ID, we also incorporate the image and text to represent an item. A fusion network is developed to merge such information for comprehensive item representations. Formally, based on embeddings of item ID $e_i^{id}$, image $e_i^{img}$ and text $e_i^{txt}$, we represent an item as follows:
\begin{align}
    \mathbf{e}_i &= \mathbf{e}_i^{id} + \mathbf{g}_{1} \odot \mathbf{e}_{i}^{img} + \mathbf{g}_{2} \odot \mathbf{e}_{i}^{txt}, \\
    \mathbf{g}_{1} &= tanh (\mathbf{W}_{1}\mathbf{m} + \mathbf{W}_{2}\mathbf{e}_{i}^{img}), \\
    \mathbf{g}_{2} &= tanh (\mathbf{W}_{3}\mathbf{m} + \mathbf{W}_{4}\mathbf{e}_{i}^{txt}), \\
    \mathbf{m} &= \mathbf{W}[\mathbf{e}_i^{id};\mathbf{e}_{i}^{img};\mathbf{e}_{i}^{txt}],
\end{align}
where $\mathbf{W} \in \mathbb{R}^{d \times 3d}$, $\mathbf{W}_{*} \in \mathbb{R}^{d \times d} $ are learnable parameters, $\odot$ is element-wise product, $ tanh(*) $ is activate function and [;] is concatenation operation. $\mathbf{m}$ integrates with various information and is used to guide the information fusing next. $\mathbf{e}_i \in \mathbb{R}^{d}$ is the comprehensive item embedding for an item $x_i$. Note that considering the intrinsic heterogeneity of these features, the gating operation is utilized in the fusion network to facilitate information fusion.

\subsubsection{Session embedding learning.}
Due to its high efficiency, self-attention is a popular choice in modeling user sequential behaviors~\cite{STAMP, SASRec, BERT4Rec}. In this work, we employ representative SASRec~\cite{SASRec} as the sequence encoder to learn session embeddings. Formally, given a session [$e_1$, $e_2$, ..., $e_m$], we obtain its embedding $\mathbf{s}  \in \mathbb{R}^{d}$ as follows:
\begin{align}
    \mathbf{s} &= {\rm seqEncoder}([\mathbf{e}_1, \mathbf{e}_2, ..., \mathbf{e}_m]),
\end{align}
where ${\rm seqEncoder}(\cdot)$ applies SASRec architecture to mine users' intents from their sequential behaviors. Note that, our \baby has no restrictions on sequence encoder, thus various models can be adopted here to handle sessions.

\subsubsection{Recommendation}
Based on learned item embedding $\mathbf{e}_i$ and session embedding $\mathbf{s}$, we can define the probability an anonymous user interacts with the item as follows,
\begin{align}
    y_i &= {\rm softmax}(\mathbf{e}_i \mathbf{s}).
\end{align}

The cross-entropy is then employed to formulate the loss of recommendation task as follows,

\begin{align}
    \mathcal{L}_{rec} = - \sum^n_{i=1} p_i \log (y_i) + (1-p_i)\log(1-y_i),
\end{align}
where $p_i$ is the ground-truth distribution of item $x_i$.

\subsection{Multi-modal Augmentation}

Considering that there are two kinds of data sparsity in SBR, \ie item-level long-tail items and session-level short sessions, we enrich data from item- and session-level simultaneously in this work.
As discussed in previous sections, due to the absence of understanding of item semantics, the existing augmentation methods for sequences~\cite{Zhou@CIKM2020, Du@CIKM2022, Xie@ICDE2022} are unable to ensure the semantic consistency of augmented views, impeding effective data enrichment. Fortunately, item multi-modal features like image and text contain rich semantic information, \eg color, size, and so on. Accordingly, we propose a multi-modal augmentation to obtain augmented views via leveraging these multi-modal features, which can fulfill robust data augmentation for items and sessions. 

\subsubsection{Item-level augmentation.} 
\begin{figure}[t]
  \centering
  \includegraphics[width=0.99\linewidth]{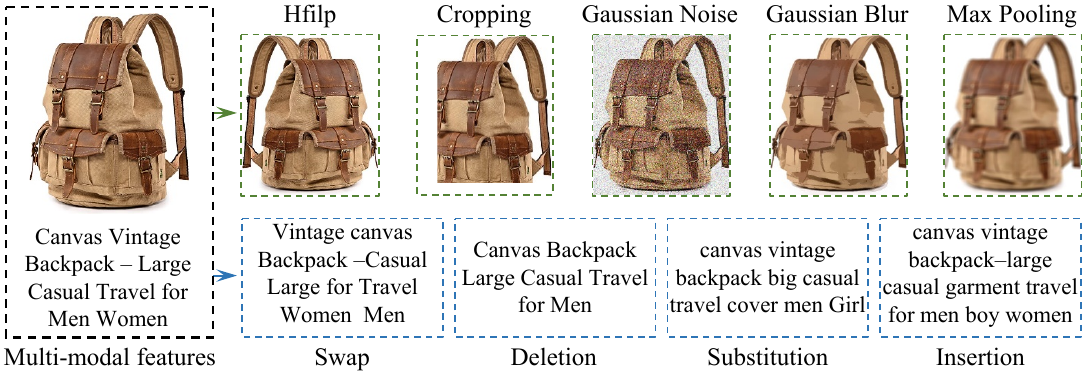}
  \caption{Multi-modal augmentation at item level. For an item, we employ augmentation techniques from CV and NLP to process its multimodal features, \eg image and text, for item-level augmentation.}\label{augmentimgtext}
\end{figure}

For an item, its semantic information is implicated in its image and text. 
Besides, some augmented techniques in Computer Vision (CV) and Natural Language Processing (NLP) can augment images and text while preserving their semantics~\cite{chen@ICML2020}.
Thus, we augment item images and text with the help of these methods which are illustrated in \fig~\ref{augmentimgtext} for convenient comprehension. 

For \textbf{images}, the following techniques are selected as augmentation methods: (1) Hflip horizontally flips images; (2) Cropping randomly removes a portion of the columns and rows of pixels at the side in images; (3) Gaussian Noise adds noise with Gaussian distribution to images; (4) Gaussian Blur blurs images with a Gaussian kernel; and (5) Max Pooling applies max pooling to handle images. 
As to description \textbf{text}, we rely on the following techniques to conduct data augmentation: (6) Swap randomly swaps some words in the text; (7) Deletion randomly deletes a few words in the text; (8) Substitution randomly substitutes words based on word similarity determined by BERT~\cite{BERT} embedding; and (9) Insertion randomly inserts words by BERT embedding similarity. Based on aforementioned augmented techniques, an augmented pool $\mathcal{A}$ is created, \ie $\mathcal{A}$ = \{ Hflip, Cropping, Gaussian Noise, Gaussian Blur, Max Pooling, Swap, Deletion, Substitution, Insertion \}. After that, for an item $x_i$, an augmented technique ${a}$ is randomly selected from $\mathcal{A}$, \ie ${a} \in \mathcal{A}$. And, the selected ${a}$ is used to obtain an augmented view $x_i^+$ of the item. According to the type of $x_i^+$, image or text, we can obtain the positive embedding $\mathbf{e}_i^+ \in \mathbb{R}^{d}$ as follows,

\begin{align}
    \mathbf{e}_i^+ = \begin{cases}
    {\rm imgEmb}(x_i^+) , & if \ x_i^+ \ is \ an \ image, \\
    {\rm txtEmb}(x_i^+), & if \ x_i^+ \ is \ text. 
    \end{cases}
\end{align}

For \textbf{item negatives}, we randomly choose $M$ items, process them with ${a}$, and obtain $M$ negative embeddings $\mathbf{e}_i^-$ = \{$\mathbf{e}_{i \cdot 1}^-$, $\mathbf{e}_{i \cdot 2}^-$, ..., $\mathbf{e}_{i \cdot M}^-$\} by $\rm {imgEmb}(\cdot)$ or ${\rm txtEmb}(\cdot)$, where $\mathbf{e}_{i \cdot k}^- \in \mathbb{R}^{d}$.
Note that, to enhance the diversity of data augmentation, different items may be augmented by distinct techniques from $\mathcal{A}$. Besides, for an item and its negatives, we employ the same augmented technique, which can ensure that the positive and negatives are located in the same space. 
In addition, we carefully select augmented techniques that can preserve an item's original semantics to build the augmented pool. As a result, the generated views hold consistent semantics, which can guarantee the acquisition of robust item augmentation and support effective item-level sparsity handling.

\subsubsection{Session-level augmentation.}
For a session $\mathcal{S}$ = [$x_1, x_2, ..., x_m$], users' intents hide behind the items they have interacted with. As shown in the top right part of \fig~\ref{macl}, based on the augmented pool $\mathcal{A}$, a technique ${c}$ is randomly selected to augment all items in $\mathcal{S}$. Accordingly, we can obtain an augmented session $\mathcal{S}^+$ = [$x_1^+, x_2^+, ..., x_m^+$], where $x_i^+$ is augmented by ${c}$. Depending on the type of $x_i^+$, \ie image or text, we represent it with $\rm {imgEmb}(\cdot)$ or $\rm {txtEmb}(\cdot)$ and get item embedding sequence [$\mathbf{e}_1^+, \mathbf{e}_2^+, ..., \mathbf{e}_m^+$]. Afterward, we input this sequence into the sequence encoder and obtain the positive session embedding $\mathbf{s}^+ \in \mathbb{R}^{d}$ as follows:
\begin{align}
    \mathbf{s}^+ &= {\rm seqEncoder}([\mathbf{e}_1^+, \mathbf{e}_2^+, ..., \mathbf{e}_m^+]).
\end{align}

As to \textbf{session negatives}, we process other sessions in the same batch with ${c}$, represent items in these generated sessions via $\rm {imgEmb}(\cdot)$ or $\rm {txtEmb}(\cdot)$, and obtain multiple negative session embeddings $\mathbf{s}^-$ = \{$\mathbf{s}_{1}^-$, $\mathbf{s}_{2}^-$, ..., $\mathbf{s}_{N-1}^-$\} by the sequence encoder, where $N$ is the batch size in model training. 
Similar to item-level augmentation, different sessions could be augmented by distinct techniques, and we employ the same augmented technique to augment a session and its negatives.
Instead of relying on dull item ID as in conventional methods~\cite{Zhou@CIKM2020, Du@CIKM2022, Xie@ICDE2022}, our multi-modal augmentation augments sessions by leveraging item multi-modal features with rich semantics. Given that every corresponding token in the anchor session and positive session holds consistent semantics, the proposed method ensures the semantic consistency of augmented sessions, contributing to tackling session-level sparsity.

\subsection{Adaptive Contrastive Loss}
We have learned positive-negative signals at both item and session levels, \ie \{ $\mathbf{e}_i$, $\mathbf{e}_i^+$, $\mathbf{e}_i^-$ \} and \{ $\mathbf{s}_i$, $\mathbf{s}_i^+$, $\mathbf{s}_i^-$ \}. As discussed before, conventional methods~\cite{Zhou@CIKM2020, Xie@ICDE2022} as in Eq.(1) are unable to distinguish signals' utility, failing to achieve effective self-supervised learning. To handle this issue, we proposed an adaptive contrastive loss to reformulate item-level contrastive loss (the same process for session level) as follows, 

\begin{align}
        \mathcal{L}^{item}_{con} &= -\sum \alpha_i \frac{<\mathbf{e}_i, \mathbf{e}_i^{+}>}{\sum_{k}<\mathbf{e}_i,  \mathbf{e}_{i \cdot k}^{-}>}, 
\end{align}
where cosine distance is used as similarity function $< , >$, and a weight $\alpha_i$ is assigned to each signal set to highlight informative signals while downplaying uninformative ones. Obviously,  $\alpha_i$ should be determined based on the utility of signal set \{ $\mathbf{e}_i$, $\mathbf{e}_i^+$, $\mathbf{e}_i^-$ \}. For efficiency and effectiveness, we apply MLP to obtain $\alpha_i$ as follows,
\begin{align}
    \alpha_i &= \text{MLP}_1([\mathbf{e}_i;\mathbf{e}_i^+;\mathbf{\bar{e}}_i^-]),
\end{align}
where $\mathbf{\bar{e}}_i^-$ = $\frac{1}{M}\sum_{k=1}^M \mathbf{e}_{i \cdot k}^-$, and $\text{MLP}_1$ is a feed-forward neural network with two hidden layers. Note that, we choose a neural network to determine $\alpha_i$ because: (1) its parameters are updated based on model performance, which can evaluate the contribution of contrastive signals on the model; (2) it can adaptively distinguish informative signals from uninformative ones based on their unique embeddings. 
Likewise, for the session level, we can implement the adaptive contrastive loss to it as follows,
\begin{align}
        \mathcal{L}^{sess}_{con} &= -\sum \beta_i \frac{<\mathbf{s}_i, \mathbf{s}_i^{+}>}{\sum_{k}<\mathbf{s}_i,  \mathbf{s}_{i \cdot k}^{-}>}, \\
        \beta_i &= \text{MLP}_2([\mathbf{s}_i;\mathbf{s}_i^+;\mathbf{\bar{s}}_i^-]).
\end{align}

At last, the whole contrastive loss is constructed to jointly handle the item- and session-level sparsity as follows,

\begin{align}
    \mathcal{L}_{con} &= \mathcal{L}^{item}_{con} + \mathcal{L}^{sess}_{con}.
\end{align}

\subsection{Training and Inferring}

\subsubsection{Training phase.}
In the training phase, to improve model performance, we optimize \baby under the joint supervision of recommendation and contrastive learning tasks as follows,
\begin{align}
    \mathcal{L} = \mathcal{L}_{rec} + \lambda \mathcal{L}_{con},
\end{align}
where the $\lambda$ is a constant controlling the strength of the contrastive learning task.

\subsubsection{Inferring phase.}
At inferring phase, given a short session produced by an anonymous user ($\mathcal{S}$ = [$x_1, x_2, ..., x_m$]), we can infer the interacted probability of all items ($y$=[$y_1$, $y_2$, ..., $y_n$]) based on item embeddings and session embedding via Eq.(10). After that, we select the top-$k$ items based on their interacted probability to form the recommendation list as follows, 

\begin{align}
    rec\_list &= [x^{'}_1, x^{'}_2, ..., x^{'}_k],
\end{align}
where $x^{'}_i$ is a recommended item with top-$k$ predicted probability.

\section{Experimental setup}
\subsection{Research Questions}
We evaluate the performance of \baby through extensive experiments with the focus to answer the following research questions:

\begin{itemize}
    \item \textbf{RQ1} Does the proposed \baby achieve better performance compared with state-of-the-art methods? (ref. Section \ref{sec:overall})
    
    \item \textbf{RQ2} What's the effect of the multi-modal augmentation? (ref. Section \ref{sec:semantic})
    
    \item \textbf{RQ3} What's the effect of the adaptive contrastive loss?  (ref. Section \ref{sec:adapter})
    
    
    \item \textbf{RQ4} How well do methods perform under data sparsity scenarios including item-level and session-level sparsity?  (ref. Section \ref{sec:itemSparsity}-\ref{sec:sessSparsity})
    
\end{itemize}

\subsection{Datasets and Preprocessing}
\begin{table}[t]
\tabcolsep 0.04in 
\centering
\caption{Statistics of all datasets.}
\begin{tabular}{cccccc}
\toprule
Datasets      & \#item  & \#session  & \#interaction & avg.length & sparsity \cr
\midrule
Cellphones	& 10,213	& 46,259	& 142,228	& 3.07	& 99.97\% \cr
Grocery	&	7,877	& 48,040	& 165,365	& 3.44	& 99.96\% \cr
Instacart	& 12,616	& 52,926	& 500,131	& 9.45	&  99.93\% \cr
\bottomrule
\end{tabular}

\label{statistics}
\end{table}

In this  work, we employ multiple public datasets to evaluate the performance of \baby and baselines including \textbf{Cellphones}, \textbf{Grocery}, and \textbf{Instacart}. The first two datasets are scratched from Amazon~\footnote{\url{http://jmcauley.ucsd.edu/data/amazon/}} and popular in SBR~\cite{SASRec, Zhou@CIKM2020, CoHHN}. Following~\cite{CoHHN}, in these datasets, user behaviors that happened within one day are formulated as a session. 
Instacart~\footnote{\url{https://www.kaggle.com/c/instacart-market-basket-analysis}}, as a dataset on the Kaggle competition, contains 3 million transactions from more than 200,000 users.  
Similar to~\cite{qin@SIGIR2021}, we randomly take 20\% transactions in Instascart as session data. Based on accessibility, the multi-modal features we incorporate in Cellphones and Grocery are item images and text, while using item text for Instacart.
For all datasets, the last item in a session is used as the ground-truth label, and the other items are utilized to mine user intents. Following ~\cite{NARM,SR-GNN}, unpopular items appearing less than 5 times and inactive sessions with a length of 1 are eliminated. Note that, we retain unpopular items in Section~\ref{sec:itemSparsity} to examine the model effect in handling item-level sparsity. Each dataset is chronologically divided into training, validation, and test sets with a ratio of 7:2:1. Statistical details of all datasets are shown in Table~\ref{statistics}. 

\subsection{Evaluation Metrics}
Following~\cite{NARM, SR-GNN, Li@WSDM2023, Yang@WWW2023}, we adopt widely used \textbf{Prec@k} (Precision) and \textbf{MRR@k} (Mean Reciprocal Rank) as our metrics to evaluate the proposed \baby and all baselines. 
In this work, the $k$ is set as 10 and 20.

\subsection{Baselines}
To examine the performance of our \baby, we employ the following two groups of competitive methods as our baselines:

\paratitle{Supervised methods:}

\begin{itemize}
    
    \item \textbf{SKNN} determines the recommended items based on the similarity between the current session and other sessions. 
    
    \item \textbf{NARM}~\cite{NARM} employs GRU with an attention mechanism to capture users' main intents. 
    
    \item \textbf{SASRec}~\cite{SASRec} uses self-attention to model user sequential behaviors. 
    
    \item \textbf{SR-GNN}~\cite{SR-GNN} views each session as a graph and applies GNN to handle session data.
    \item \textbf{UniSRec}~\cite{hou@KDD2022} explores item description text to obtain universal sequence representations.

    \item \textbf{DGNN}~\cite{Li@WSDM2023} exploits explicit and implicit item relations with GNN to improve user intent understanding.
    
\end{itemize}

\paratitle{Self-supervised methods:}
\begin{itemize}
    
    \item \textbf{COTREC}~\cite{Xia@CIKM2021} relies on Relation Mapping to augment sessions, where session embeddings learned from item and session graphs are viewed as positives.

    \item \textbf{CL4SRec}~\cite{Xie@ICDE2022} augments user sequences via Crop, Mask and Reorder.

    \item \textbf{DuoRec}~\cite{Qiu@WSDM2022} applies Dropout and Retrieval for sequence augmentation and investigates the problem of representation degeneration.

    \item \textbf{ICLRec}~\cite{chen@WWW2022} clusters user intents and views the similar (dissimilar) intents as positives (negatives).

\end{itemize}

\begin{table*}[t]
\tabcolsep 0.03in 
    \caption{Performance comparison of \baby with baselines over four datasets. The $\spadesuit$ denotes contrastive learning-based methods. The results (\%) produced by the best baseline and the best performer in each row are underlined and boldfaced respectively. Significant improvements of \baby over the best baseline (*) is determined by the t-test ($p < 0.01$).}
    \begin{tabular}{c cccc cccc cccc}  
    \toprule  
    \multirow{2}{*}{Method}& 
    \multicolumn{4}{c}{Cellphones}&\multicolumn{4}{c}{Grocery}&\multicolumn{4}{c}{Instacart}\cr      
    \cmidrule(lr){2-5} \cmidrule(lr){6-9} \cmidrule(lr){10-13}
    \cmidrule(lr){2-3} \cmidrule(lr){4-5} \cmidrule(lr){6-7}    \cmidrule(lr){8-9}  \cmidrule(lr){10-11}    \cmidrule(lr){12-13} 
    &Prec@10&MRR@10&Prec@20&MRR@20  &Prec@10&MRR@10&Prec@20&MRR@20  &Prec@10&MRR@10&Prec@20&MRR@20\cr  
    \midrule  
    SKNN	&9.33&8.02&12.84&8.24	&41.70&30.20&43.90&30.35	&4.45&1.36&7.75&1.65\cr  
    NARM	&12.92&9.53&14.36&9.66	&46.88&41.92&48.19&42.11	&5.24&1.85&8.50&2.07\cr  
    SASRec	&15.65&8.92&17.92&9.08	&43.88&31.39&46.52&31.57	&5.87&2.31&9.08&2.51\cr  
    SR-GNN	&12.08&9.43&13.72&9.60	&45.60&41.14&47.44&41.28	&6.01&2.08&9.04&2.28\cr  
    UniSRec	&15.27&9.69&18.09&\underline{9.96}	&47.09&34.32&49.31&34.48	&6.02&2.19&8.74&2.37\cr  
    DGNN	&15.21&\underline{9.73}&17.22&9.86	&47.21&\underline{42.18}&48.92&\underline{42.29}	&6.23&2.27&9.40&2.51\cr
    COTREC$^\spadesuit$	&\underline{15.96}&9.49&\underline{18.53}&9.80	&\underline{47.30}&33.58&\underline{49.42}&33.77	&6.12&1.79&9.69&2.04\cr
    CL4SRec$^\spadesuit$	&14.51&9.59&17.18&9.70	&44.99&38.19&46.77&38.32	&\underline{6.64}&\underline{2.51}&\underline{9.97}&\underline{2.76}\cr  
	DuoRec$^\spadesuit$	 &14.45&9.40&17.48&9.62	&44.57&38.58&46.62&38.75	&6.37&2.48&9.82&2.61\cr
	ICLRec$^\spadesuit$	&13.49&8.33&16.01&8.68	&41.49&36.67&43.52&36.89	&6.11&2.46&9.44&2.58\cr
	\midrule
	\baby	&{$ \bf 16.62^*$}&{$ \bf 12.39^*$}&{$ \bf 19.46^*$}&{$ \bf 12.53^*$}	&{$ \bf 48.04^*$}&{ $ \bf 43.09^*$ }&{ $ \bf 50.25^*$ }&{ $ \bf 43.24^*$ }	&{ $ \bf 7.43^*$ }&{ $ \bf 2.96^*$ }&{ $ \bf 10.55^*$ }&{ $ \bf 3.19^*$ }\cr
	$improv.$	&4.14\%&27.34\%&5.02\%&25.80\%	&1.56\%&2.16\%&1.68\%&2.25\%	&11.90\%&17.93\%&5.82\%&15.58\%\cr
	\bottomrule
    \end{tabular}
    \label{performance}
\end{table*}  

\subsection{Implementation Details}
To ensure a fair comparison, hyperparameters of \baby and all baselines are determined via grid search according to their performance on Prec@20 in the validation set. For \baby,  we fix embedding size $d$ = 100, pick balance coefficient $\lambda$ = 0.01, choose $M$ = 100 negatives for item-level contrastive learning, and set mini-batch size $N$ = 100. The \baby is optimized via Adam with the initial learning rate at $0.001$.
Given that the output dimension of GoogLeNet and BERT are 1024 and 768 respectively, we utilize the PCA algorithm to reduce them to 100.
The source codes are available online\footnote{\url{https://github.com/Zhang-xiaokun/MACL}}.

\section{Results and Analysis}

\subsection{Overall Performance (RQ1)}\label{sec:overall}

The overall performance of \baby and all baselines is reported in Table~\ref{performance},  where we can obtain the following observations:

Firstly, the performance of baseline methods varies significantly depending on contexts (datasets). Taking CL4SRec as an example, it performs well in Instacart but poorly on the other three datasets. It indicates that SBR is indeed a challenging task that requires revealing user intents within limited information. Due to their inability to adequately handle the data sparsity caused by long-tail items and short sessions in SBR, these methods are unable to consistently deliver high performance across various contexts.

Secondly, among supervised methods, UniSRec and DGNN achieve impressive performance. We contend that the effectiveness of UniSRec benefits from its incorporation of item description text. The item text contains its specific semantic information, such as size and brand, which contributes to user preferences understanding. As to DGNN, it builds two graph structures by exploiting explicit and implicit item relations and utilizes GNN to learn node embeddings on these graphs. Similar to Relation Mapping, such a method enriches session data to some extent, enabling competitive performance in some instances.

Thirdly, in self-supervised baselines, COTREC outperforms other methods with a large margin on the Prec@k metric. Instead of simply disturbing the item ID sequence like Crop, Mask, and Reorder, COTREC resorts to Relation Mapping for session augmentation. This operation preserves session information and improves performance. However, the performance of COTREC on MRR@k is far from satisfactory. It implies that the target item frequently appears at the end of the list, which decreases the usefulness of COTREC in practical situations. Moreover, CL4SRec, DuoRec and ICLRec produce competitive results in Instacart, while their performance on other datasets is poor. Referring to Table~\ref{statistics}, sessions in Instacart are much longer than their counterparts in other datasets. It indicates that commonly used augmentations have intrinsic inadequacy (ref. Section~\ref{sec:revisit}) in solving session-level sparsity, and prevent models using them from performing well in short sessions. 

Finally, the proposed \baby achieves consistent superiority over all baselines in terms of all evaluation metrics on all datasets, which confirms its effectiveness for SBR. In particular, \baby surpasses the best baselines in Prec@20 and MRR@20 by 5.02\% and 25.80\% on Cellphones, 1.68\% and 2.25\% on Grocery, and 5.82\% and 15.58\% on Instacart. The \baby jointly handles item-level sparsity and session-level sparsity existing in SBR, which contributes to its strong performance. Instead of relying on dull item ID, the multi-modal augmentation in \baby leverages item multi-modal features to achieve robust data augmentation. Besides, an adaptive contrastive loss is also presented in \baby to distinguish informative contrastive signals from uninformative ones. These novel techniques are capable of bridging the gaps left by current solutions, thus enabling \baby to achieve effective self-supervised learning and fulfill the satisfactory intent prediction. Furthermore, our \baby obtains inspiring improvements over baselines in terms of MRR@k. This makes it easy for target items to be discovered by potential users, which obviously can enhance user experience and encourage online consumption.

\subsection{The effect of multi-modal augmentation (RQ2)}\label{sec:semantic}

\begin{figure}[t]
  \centering
  \includegraphics[width=0.9\linewidth]{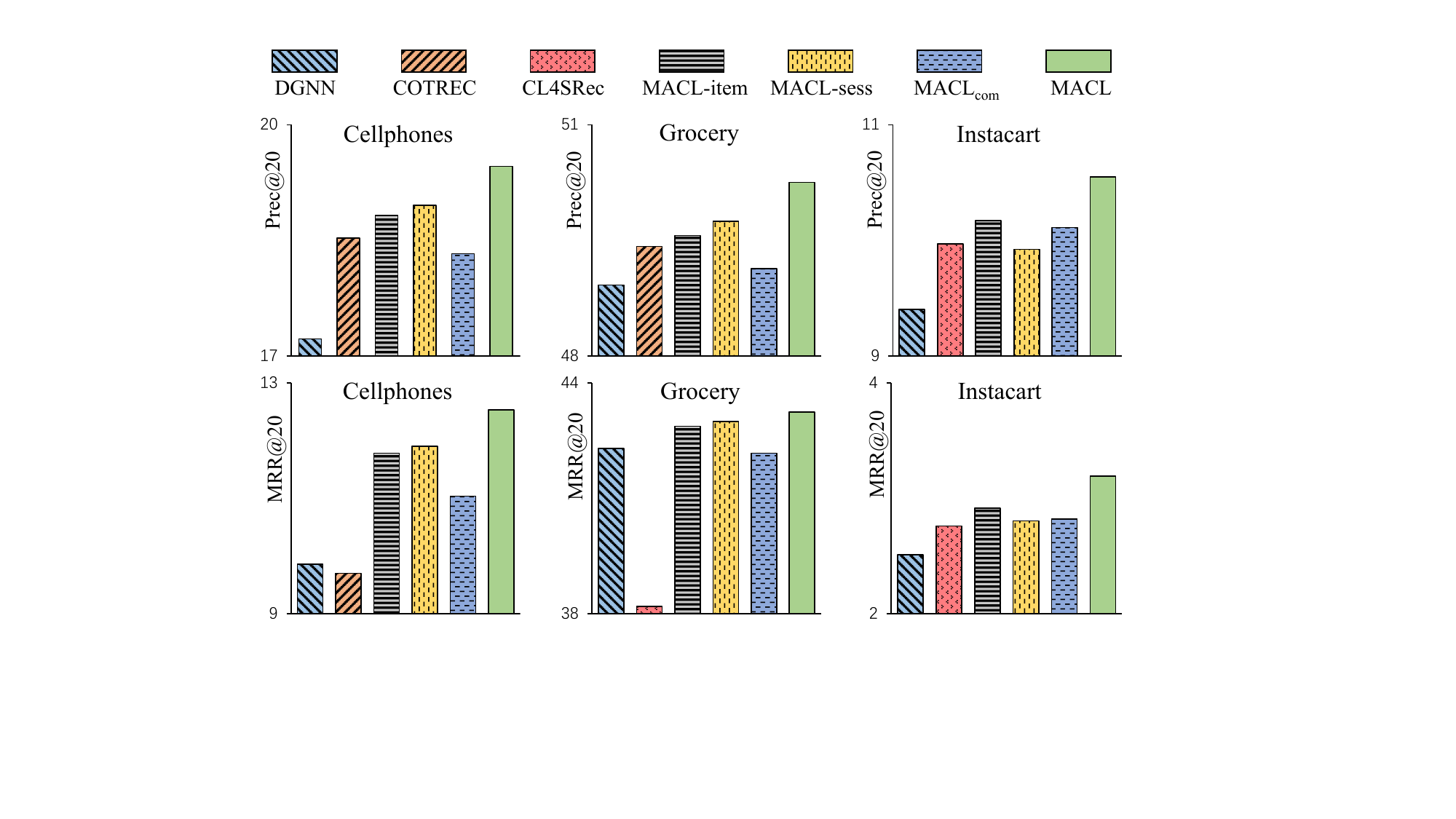}
  \caption{The effect of multi-modal augmentation.}\label{sematics}
\end{figure}

The key novelty of the proposed \baby is that it devises a multi-modal augmentation to obtain robust data augmentation at both item and session levels via leveraging item multi-modal features. In this part, we build the following variants of \baby to demonstrate the efficacy of the proposed technique:
\baby-item and \baby-sess remove item-level and session-level contrastive learning from \baby, respectively, that is, they only consider one aspect of data sparsity; \babyx$_{com}$  applies commonly used methods to augment items and sessions without leveraging item multi-modal features, where Dropout is used for item augmentation and Crop, Mask, as well as Reorder, are utilized to augment sessions.

We can obtain the following insights from \fig~\ref{sematics}: (1) \baby defeats both \baby-item and \baby-sess, which indicates that there are two sparsity issues in SBR and considering only one of them can not achieve satisfactory performance. In contrast, jointly coping with these issues, the proposed \baby can effectively improve SBR. (2) In most instances, \baby-sess outperforms \baby-item. In other words, compared with session-level sparsity, handling item-level sparsity can bring larger improvements for SBR. It suggests that the item-level sparsity issue plays  a dominant role in SBR. Unfortunately, existing methods pay little attention to this issue, which seriously limits their performance; (3) \baby performs much better than \babyx$_{com}$. It demonstrates the effectiveness of the proposed multi-modal augmentation. Leveraging item multi-modal features with rich semantics, the multi-modal augmentation is capable to generate augmented views holding consistent semantics, enabling high-quality data augmentation; (4) \baby-item achieves better performance than COTREC and CL4SRec. All these three methods focus on session-level sparsity. The results prove that conducting data augmentation on multi-modal features is a promising antidote to address data sparsity in SBR. It demonstrates the rationality and effectiveness of the proposed multi-modal augmentation again.

\subsection{The effect of adaptive contrastive loss (RQ3)}\label{sec:adapter}

As discussed before, distinct positive-negative signals possess various utility for model performance. Therefore, we present a novel adaptive contrastive loss to reformulate contrastive loss, where informative signals are emphasized while uninformative ones are  downplayed. \baby-adp removes the adaptive contrastive loss from \baby, \ie it directly applies Eq.(1) to conduct item-and session-level contrastive learning. As presented in Table~\ref{adpater}, \baby achieves significant superiority over \baby-adp in all cases, which indicates the effectiveness of adaptive contrastive loss. We believe that the adaptive contrastive loss evaluates the signals' contribution to model performance, reassigns their weights accordingly, and finally improves self-supervised learning for SBR. Moreover, \baby-adp outperforms other competitive baselines in most instances. This serves as more evidence for the necessity and effectiveness of utilizing multi-modal features for data augmentation. 

\begin{table}[t]
\centering
\small
\tabcolsep 0.02in 
    \caption{The effect of adaptive contrastive loss.}
    \begin{tabular}{c c ccccc}  
    \toprule 
     Datasets   &Metrics    &DGNN &COTREC$^\spadesuit$     &CL4SRec$^\spadesuit$ &\baby-adp  &\baby    \cr
    \midrule  
    \multirow{2}*{Cellphones}   
    &Prec@20 &17.22&18.53&17.18&18.97&{$ \bf 19.46^*$} \cr
    &MRR@20 &9.86&9.80&9.70&12.04&{$ \bf 12.53^*$} \cr
    \midrule
    \multirow{2}*{Grocery}
    &Prec@20 &48.92&49.42&46.77&49.84&{ $ \bf 50.25^*$ } \cr
    &MRR@20 &42.29&33.77&38.32&42.80&{ $ \bf 43.24^*$ }  \cr
    \midrule
    \multirow{2}*{Instacart}
    &Prec@20 &9.40&9.69&9.97&9.89&{ $ \bf 10.55^*$ } \cr
    &MRR@20 &2.51&2.04&2.76&2.85&{ $ \bf 3.19^*$ }  \cr
	\bottomrule
 
    \end{tabular}
    \label{adpater}
\end{table}  

\begin{table}[tp]
\small
\tabcolsep 0.03in 
  \centering
  \caption{Runtime (seconds) of \baby and \baby-adp during training and inference.}  
  \label{runtime}  
    \begin{tabular}{c ccc ccc}  
    \toprule
    \multirow{2}{*}{Method}&\multicolumn{2}{c}{Cellphones}&\multicolumn{2}{c}{Grocery}&\multicolumn{2}{c}{Instacart}\cr      
    \cmidrule(lr){2-3} \cmidrule(lr){4-5} \cmidrule(lr){6-7}
    &{Training}&{Inference}&{Training}&{Inference}&{Training}&{Inference}  
    \cr  
    \midrule  
    \baby-adp   &205.69&11.91&184.77 &10.26&105.34&8.00\cr  
    \baby       &207.83&12.24&186.52 &10.84&109.11&8.31\cr
	\bottomrule
    \end{tabular}  
\end{table} 

To further examine the efficiency of the proposed adaptive contrastive loss, we record the runtime of \baby and \baby-adp across all datasets. Specifically, we run both \baby and \baby-adp on a single ‌NVIDIA GeForce RTX 4090 GPU server with 24GB memory, and record their average runtime per epoch during training and inference phases.
As shown in Table~\ref{runtime}, the runtime increase due to the adaptive contrastive loss in MACL is minimal, with only about 2 seconds in the training phase and 0.3 seconds during inference. We attribute this small increase in runtime to the utilization of a simple yet effective MLP structure in the adaptive contrastive loss. Given that this modest increase in runtime is associated with a noticeable boost in accuracy, it suggests that the adaptive contrastive loss is essential for enhancing the model’s performance without significantly compromising efficiency. This can be interpreted as evidence of the cost-effectiveness of the adaptive contrastive loss, making it a valuable addition for improving model effectiveness in practice.

\subsection{Handling item-level sparsity (RQ4)}\label{sec:itemSparsity}

\begin{figure}[t]
  \centering
  \includegraphics[width=0.9\linewidth]{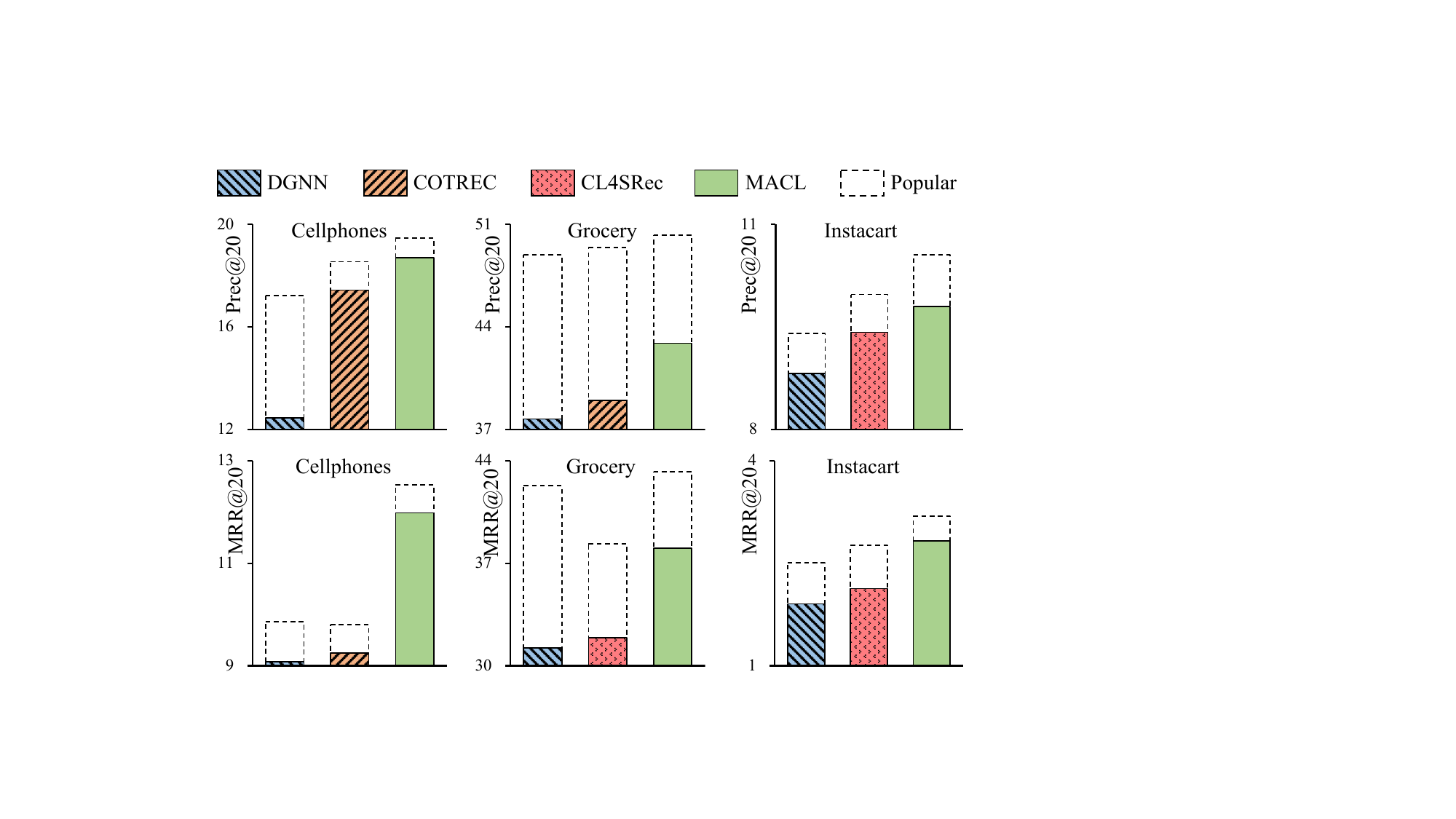}
  \caption{Performance under item-level sparsity.}\label{itemSparsity}
\end{figure}

Item-level sparsity, \ie long-tail items, extensively exists in SBR scenarios. However, existing methods always blindly delete unpopular items in their experiments~\cite{NARM, Xia@CIKM2021, Xie@ICDE2022, Li@WSDM2023}. In this part, we retain unpopular items to examine the performance of \baby and competitive baselines under genuine SBR scenarios, where the statistics are summarized in Table~\ref{statisticsSpa}. We can draw the following conclusions from \fig~\ref{itemSparsity}: 
(1) All models exhibit a declined performance when dealing with unpopular items, demonstrating the necessity and difficulty of handling long-tail items in SBR; (2) Among these methods, our \baby has the least performance degradation under unpopular items. It indicates the effectiveness of \baby in tackling long-tail items. Benefiting from enriching item data with multi-modal augmentation and explicitly handling item-level sparsity by item-level contrastive learning, \baby can gauge user interest in long-tail items and provide customized services; (3) \baby presents overwhelming advantages over other competitive baselines under long-tail situations. It demonstrates the effectiveness of \baby in addressing item-level sparsity in SBR once more. We believe that this merit of \baby contributes to satisfying users in real scenarios, thereby promoting the development of SBR.

\begin{table}[t]
\small
\tabcolsep 0.03in 
\centering
\caption{Statistics of all datasets under unpopular items.}
\begin{tabular}{cccccc}
\toprule
Datasets      & \#item  & \#session  & \#interaction & avg.length & sparsity \cr
\midrule
Cellphones+	& 14,534(+4,321)	& 53,238	& 164,532	& 3.09	& 99.98\% \cr
Grocery+	&	18,499(+10,622)	& 55,380	& 198,549	& 3.59	& 99.98\% \cr
Instacart+	& 21,432(+8,816)	& 53,340	& 519,389	& 9.74	&  99.95\% \cr
\bottomrule
\end{tabular}

\label{statisticsSpa}
\end{table}

\subsection{Handling session-level sparsity (RQ4)}\label{sec:sessSparsity}

\begin{figure}[t]
  \centering
  \includegraphics[width=0.95\linewidth]{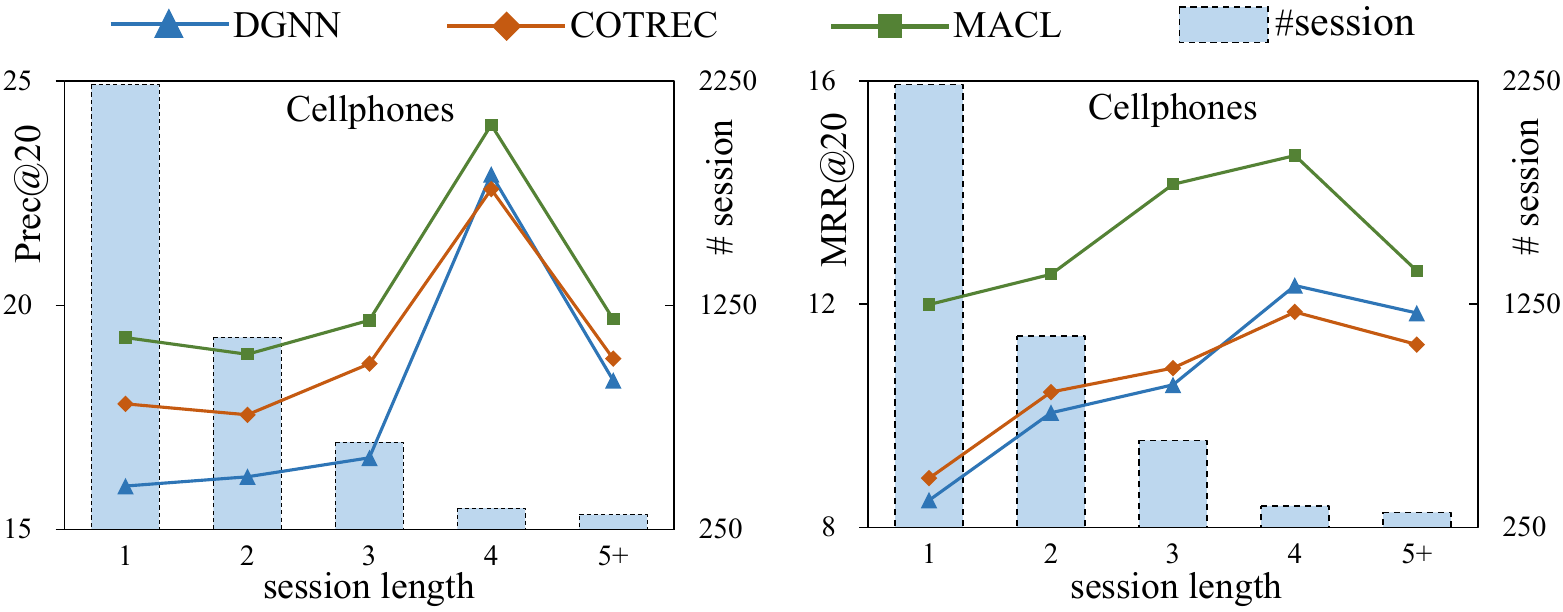}
  \caption{Performance under session-level sparsity.}\label{sessSparsity}
\end{figure}

As another intrinsic issue that SBR suffers from, session-level sparsity, \ie short sessions, can seriously deteriorate recommendation performance. In this section, we examine the performance of \baby and representative baselines under various session lengths. We present the performance patterns in representative Cellphones, and similar observation can be obtained on the other datasets. The results are plotted in  \fig~\ref{sessSparsity}, where the following insights are noted: (1) Short sessions make up a large proportion in SBR, which makes it hard to accurately predict user intents with such limited behaviors. It indicates the necessity to handle session-level sparsity for effective SBR. (2) Compared with long sessions, all models fail to achieve good performance under short sessions. It is straightforward since that there is not enough information to model user behaviors in short sessions. Fortunately, contrastive learning could be a remedy for this issue. For instance, on MRR@20 in Cellphones, COTREC is defeated by DGNN under long sessions, whereas it outperforms DGNN in short ones. We argue that session augmentation used in contrastive learning based methods can enrich session data, thus leading to their impressing performance in short sessions; (3) The proposed \baby accomplishes greater gains over baselines on short sessions than on long ones. It demonstrates the effectiveness of \baby in addressing session-level sparsity. By leveraging item multi-modal features instead of dull item ID, the multi-modal augmentation devised in \baby can obtain robust data augmentation, contributing to handling short sessions; (4) Our \baby performs better than competitive baselines at all session lengths in all metrics. It verifies the superiority of \baby in improving SBR again. 


\section{Related Work}

\subsection{Session-based Recommendation}
With their strong representation capability, neural networks have gained predominance for SBR recently~\cite{DIDN, DIMO}. Many works introduce various neural architectures into SBR to mine user intents such as RNN~\cite{GRU4Rec,NARM}, Self-attention~\cite{DPAN4Rec,BERT4Rec,SASRec} and GNN~\cite{SR-GNN,Wang@PR2024}. To enrich training data, some methods exploit extra sessions~\cite{Song@WWW2023}, multiple user intents~\cite{Zhang@WSDM2023}, and multi-level relations among items~\cite{Li@WSDM2023}. Moreover, there are some methods leveraging side information to facilitate user preferences capturing like categories~\cite{cai@SIGIR2021,Hao@PR2023}, brands~\cite{song@CIKM2021}, price~\cite{CoHHN, BiPNet} and text~\cite{hou@KDD2022, FineRec}. Although greatly boosting the development of SBR, unfortunately, existing methods are still unable to handle data sparsity in SBR effectively.

\subsection{Contrastive Learning}
Contrastive learning aims to improve representation learning via maximizing agreement between augmented views of an instance~\cite{chen@ICML2020}. Due to its ability in tackling session-level data sparsity, CL is prevailing in sequence modeling recently. As the core of CL, data augmentations (ref. Section~\ref{sec:revisit} for detail) have been widely explored in the task such as item ID sequence perturbation (Crop, Mask and Reorder)~\cite{Zhou@CIKM2020, Xie@ICDE2022, Du@CIKM2022, Liu@CoRR}, Dropout~\cite{Qiu@WSDM2022}, Retrieval~\cite{Qiu@WSDM2022, LXW@WSDM2023, WangL@CIKM2022,Cui@CoRR2025} and Relation Mapping~\cite{DHCN, Xia@CIKM2021, WZY@CIKM2022}. 
However, they mostly conduct augmentation based on item ID while unable to explore item multi-modal features with rich semantics, which results in their failure in generating semantically consistent augmented views. In addition, they fail to distinguish the various utility of different positive-negative signals, which prevents them from engaging in effective self-supervised learning.

\section{Conclusion and Future Work}

\subsection{Theoretical and Practical Implications}
In this study, we scrutinize contrastive learning based methods in session-based recommendation and spot their main limitations including inability to handle item-level sparsity, corrupt session augmentation and failure in distinguishing varying utility of contrastive signals. Accordingly, we propose a novel multi-modal adaptive contrastive learning framework (\baby). Specifically, a multi-modal augmentation is devised to augment both items and sessions. Instead of relying on dull item IDs, the proposed multi-modal augmentation leverages item multi-modal features to obtain augmented views with consistent semantics. Moreover, we present an adaptive contrastive loss to reformulate contrastive loss, where informative signals are emphasized while uninformative ones are downplayed for effective self-supervised learning. 

Extensive experiments on multiple public datasets demonstrate the superiority of \baby over state-of-the-art methods of SBR. Further analysis also indicates the effectiveness of \baby in handling data sparsity issues including long-tail items and short sessions in SBR. In addition, our study finds that: (1) handling both item- and session-level sparsity contributes to improving SBR; (2) relying on item multi-modal features instead of item IDs helps to build robust augmentations; (3) highlighting varying utility of contrastive signals can achieve effective self-supervised learning. These insights expand our understanding for contrastive learning in session-based recommendation and offer valuable reference points for future research.

\subsection{Limitation and Future Work}
Although \baby significantly improves contrastive learning in session-based recommendation, it still faces several issues that should be addressed in future work. On one hand, \baby currently focuses only on text and images for data augmentation, overlooking other available item information, such as categories, brands, and reviews. Therefore, we plan to incorporate these rich item features to create fine-grained item representations and provide diverse alternatives for data augmentation. On the other hand, we adhere to the conventional and straightforward method (random operation) to determine the negative samples. However, random-like selection for negatives may decrease the quality of contrastive signals and impede effective self-supervised learning. Consequently, a promising direction for future work is to focus on sampling more informative negatives to further enhance self-supervised learning in SBR.

\section{Acknowledgment}
This work is supported by the Natural Science Foundation of China with the grant ID of (No.62376051, No.62076046), the Fundamental Research Funds for the Central Universities award number (DUT24LAB123), and the Liaoning Provincial Natural Science Foundation Joint Fund Program(2023-MSBA-003). 



\bibliographystyle{ACM-Reference-Format}
\bibliography{ref}

\end{document}